\documentclass[10pt,conference,twocolumn]{IEEEtran}

\usepackage{cite}
\usepackage{balance}
\usepackage{color}
\usepackage{epsfig}
\usepackage{epstopdf}
\usepackage{graphicx}
\usepackage{amssymb,amsmath}
\usepackage{bm}
\usepackage{bbm}
\usepackage{amsthm}
\usepackage{placeins}
\usepackage{lipsum}
\usepackage{algpseudocode}
\usepackage{algorithm}

\ifCLASSOPTIONcompsoc
\usepackage[caption=false,font=footnotesize,labelfont=sf,textfont=sf]{subfig} 
\else
\usepackage[caption=false,font=footnotesize]{subfig}
\fi

\newcommand{\comment}[1]{}

\usepackage{textcomp}
\usepackage{xcolor}
\def\BibTeX{{\rm B\kern-.05em{\sc i\kern-.025em b}\kern-.08em
		T\kern-.1667em\lower.7ex\hbox{E}\kern-.125emX}}

\begin{document}	
\title{ Underlaid FD D2D Communications in Massive MIMO Systems via Joint Beamforming and Power Allocation }

\author{\normalsize
	Hung V. Vu and Tho Le-Ngoc\\
	\IEEEauthorblockA{
		Department of Electrical \& Computer Engineering, McGill University, Montr\'{e}al, QC, Canada \\
		Emails: hung.vu2@mail.mcgill.ca, tho.le-ngoc@mcgill.ca} 
}
%
%
	\maketitle 


\begin{abstract}	
	This paper studies the benefits of incorporating underlaid full-duplex (FD) device-to-device (D2D) communications into massive multiple-input-multiple-output (MIMO) \emph{downlink} systems. Due to the nature of cellular downlink and FD D2D transmission, the performances of cellular and D2D services are severely impaired due to high interference caused by base-stations (BSs) and D2D transceivers. As a consequence, integrating a large number of D2D links into exiting cellular networks might degrade the system performances. To overcome this challenge, utilizing the large uniform linear array (ULA) equipped at BSs, we propose a joint beamforming and power allocation design for average sum-rate maximization while considering the effects of interference to both cellular and D2D transmission. The problem formulation leads to a nonconvex vector-variable optimization problem, where we develop an efficient solution using a fractional programming (FP) based approach. Numerical results show that, at sufficiently high self-interference cancellation (SIC) levels and numbers of active D2D links, the FD D2D transmission provides a significant sum-rate improvement as compared to the half-duplex (HD) counterpart and pure cellular systems in absence of D2D. 	
\end{abstract}

 \vspace{-1ex}

\section{Introduction}\label{sec:intro}
In recent years, device-to-device (D2D) communications has emerged as an innovative technology for future cellular networks \cite{JAndrewComMag}. Instead of traversing through base-station (BS), D2D communications enables the cellular users to communicate directly with each others, thereby reducing network congestion, shortening packet delay and enhancing spectral efficiency. 
In cellular networks, separate sets of carrier frequencies are allocated to the uplink and downlink transmission. The underlaid D2D services generally favors the use of uplink spectrum \cite{JAndrewComMag}. The main reason is that, in downlink transmission, the cellular receivers are typically located close to the D2D transmitters, thus reusing the downlink spectrum can substantially increase the D2D-to-cellular interference. Moreover, high transmit power level at the BS can also cause severe interference to the D2D receivers. Therefore, interference control is crucial to enable the underlaid D2D services in cellular downlink transmission.

%

In cellular systems, the employment of massive multiple-input-multiple-output
(MIMO) equipped BSs in conjunction with coordinated resource allocation between cellular and D2D transmitters is an appealing solution to address this challenge. 
With a sufficiently large number of antennas and intelligent beamformer design, the BS is capable of forming very narrow beams aiming toward the intended cellular receivers, thereby resulting in extremely low interference to co-channeled cellular/D2D users. 
Meanwhile, implementing resource allocation strategies at D2D transmitters allow to effectively mitigate the interference caused by the D2D transmission at both cellular and D2D receivers. In this paper, we shall develop a joint beamforming and power allocation algorithm in massive MIMO muti-cell systems being underlaid by D2D transmission for network throughput maximization, while ensuring the quality of service (QoS) for both cellular and D2D users. \emph{Our focus is on the downlink where time-frequency resources are fully reused by the underlaid D2D transmission. }

In existing research, there has been considerable interest in designing the precoding/beamforming and/or power allocation techniques for massive MIMO cellular networks being underlaid by D2D services. 
For instance, power control at both cellular and D2D users with fixed beamformers was investigated in \cite{Ghazan_MC} for multi-cell massive MIMO systems with underlaid D2D in order to maximize the minimum spectral efficiency (SE). 
For single-cell massive MIMO cellular system being underlaid by D2D users, Chen \textit{et al.} in \cite{Chen_SC} proposed a simple rate adaptation method based on stochastic geometry approach to minimize the interference to cellular users. 
Rate adaptation based on a stochastic geometry approach was extended for the multi-cell setting in \cite{He_SG}. For underlaid D2D systems in absent of cellular transmission, Shen \textit{et al.} adopted the matrix fractional programming techniques, solving the coordinated joint scheduling, power control, and beamforming so as to optimize the network sum-rate \cite{ShenFRD2D}. In this system, each D2D link was equipped with single-user (SU) MIMO transmission.
In prior works, D2D studies have developed and evaluated under the consideration of half-duplex (HD) D2D communications, where a D2D user can either transmit or receive on a single channel, but not simultaneously. Given a number of encouraging FD designs, the integration of FD in D2D communications is an attractive solution for the development of new architectures and algorithms in cellular networks.  
As the FD D2D nodes use a single carrier frequency in both transmitting and receiving signals, the use of FD D2D allows to alleviate the frequency resource demand for underlaid D2D transmission. 
Recently, the power control problems were investigated for underlaid FD D2D in \cite{FDD2DAli, FDD2DHuang, HungTVT19}. 
Under this line of works, single antenna transmission was adopted at both D2D and cellular users. 

This paper focuses on exploring impact of incorporating underlaid FD D2D transmission into existing multi-cell networks where BSs are equipped with massive MIMO transmission. Utilizing the uniform linear array (ULA) antennas at BSs, our objective is to design a joint beamforming and power allocation algorithm so as to alleviate the interference between cellular and D2D transmission and optimize the overall network sum-rate. The proposed algorithm is based on the fractional programming (FP) approach, which is developed to solve general noncovex optimization problem in which the objective function is in form of the sum-functions-of-ratio. 
In wireless communications systems, FP has recently been employed extensively as a low-complexity solution to solve the energy and spectral efficiency optimization problems (e.g., \cite{FRCheung, FRZappone, FPWu, FPWuPart2, ShenFRD2D} and the references therein). The FP-based approach
	takes advantage of the fractional structure of
	the nonconvex optimization problem by directly looking at
	the objective function decomposition. 
	Specifically, the FP-based algorithm approximates the nonconvex optimization problem by applying a quadratic transformation to the fractional argument terms of the objective function to ensure that the transformed objective function is concave. 
	While our algorithm utilizes the FP theory developed in \cite{FPWu} applied to separate power control and beamforming design in cellular systems, the extension to joint beamforming and power allocation in underlaid D2D cellular networks is not trivial and requires new derivations.
	
Based on the developed FP-based algorithm, we study the average network sum-rates per cell under the impact of important network parameters including SIC levels and number of active D2D links in each cell. 	
	We adopt the network sum-rates of HD D2D cellular systems and pure cellular systems (in absence of D2D), both also employing the FP-based algorithm, as the benchmarks to demonstrate the advantages of FD D2D transmission. An interesting observation is that adopting the FP-based algorithm to FD D2D cellular systems significantly improves the achievable network throughput (over a cell) and vastly outperforms that achieved by the HD D2D counterpart and pure cellular systems, but only under sufficiently high SIC levels and number of active D2D links. 
	

\section{System Model and Problem Formulation}
\subsection{System Model} 
\begin{figure} [htb!]
	\centering
	\includegraphics[scale=0.48]{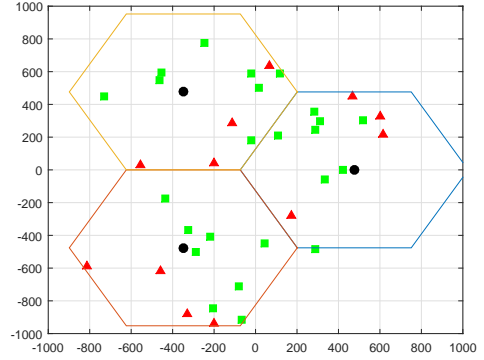}
	\caption{A underlaid D2D cellular network with multi-cell setting (black circle: BS, red triangle: cellular user, green square: D2D transceiver). For a clear representation, we only plot one D2D transceiver per link.} 
	\label{Fig:SystemModel}
\end{figure}  
As illustrated in Fig. \ref{Fig:SystemModel},
we consider a hybrid network
including cellular downlinks and D2D links operating in HD and FD modes, respectively. The network
consists of $B$ base-stations (BSs) located in multiple hexagonal cells. 
We assume that both multiple cellular and D2D transceivers are randomly located within the cell region. The associated receiver with a D2D transceiver is located at a fixed distance away with isotropic direction. This paper considers single antenna transmission at all cellular/D2D users, while each BS is equipped with $A$ antennas. Each BS is capable of serving $M$ cellular users simultaneously in a time-frequency slot ($A \ge M$). In addition, the downlink resources are fully reused by multiple FD D2D links, and the universal frequency reuse scheme (UFR) is employed. 
The sets of cellular users served by the BS $b$ are denoted as $\mathcal{C}_b,\, b=1,\dots,B$. Meanwhile, let $\mathcal{D}$ be the collection of all D2D transceivers in the network;  
and $N = \lvert \mathcal{D} \rvert/(2B)$ is the number of D2D links per cell. 
The transmit power vector of D2D users is defined as  $\mathbf{p} =[P_1, \dots, P_{2BN}] \in \mathbb{R}_{+}^{2BN}$, where each element $P_n$ denotes the transmit power at D2D transceiver $n$. The path-loss $L$ is computed as $L = C r^{-\alpha}$ where $r$ is the distance, $\alpha > 2$ is the path-loss exponent, and $C$ denotes the reference path-loss.  
Furthermore, the system uses the time division multiplexing (TDD); we assume channel reciprocity, i.e., the downlink channel is Hermitian transpose of the uplink channel.

Given the system model, the received signals at cellular user $m$ in cell $b$ is given by 
\begin{align} \label{eq:signal:cell} \small
	y^{(\text{c})}_{bm} &= L^{(\text{cc})}_{b, bm} \left(\mathbf{h}^{(\text{cc})}_{b,bm} \right)^{\mathsf{H}} \mathbf{v}_{bm} x^{(\text{c})}_{bm} \nonumber \\
	&+  \underbrace{\sum\limits_{j\in \mathcal{C}_b\setminus \{m\}}  L^{(\text{cc})}_{b, bm} \left(\mathbf{h}^{(\text{cc})}_{b,bm} \right)^{\mathsf{H}}  \mathbf{v}_{bj} x^{(\text{c})}_{bj}  }_\textrm{Intra-cell cellular interference}  \nonumber \\
	&+  \underbrace{\sum\limits_{i \neq b } \sum\limits_{j\in \mathcal{C}_i}  L^{(\text{cc})}_{i,bm} \left(\mathbf{h}^{(\text{cc})}_{i,bm} \right)^{\mathsf{H}}  \mathbf{v}_{ij} x^{(\text{c})}_{ij}  }_\textrm{Inter-cell cellular interference}  \nonumber \\
	&+ \underbrace{ \sum\limits_{j\in \mathcal{D}} L^{(\text{dc})}_{j,bm} {h}^{(\text{dc})}_{j,bm} x^{(\text{d})}_j  }_\textrm{D2D-to-cellular interference}  + z^{(\text{c})}_{bm} ,
\end{align}
where $\mathbf{h}^{(\text{cc})}_{l, bm}  \in \mathbb{C}^A$ ($L^{(\text{cc})}_{l, bm} \in \mathbb{R}_+$), $l\in \{b,i\},$ denotes the channel fading vector (path-loss) from for cellular user $m$ in cell $b$ to the BS $l$. 
Correspondingly, $\mathbf{v}_{uv} \in \mathbb{C}^A, \, uv\in\{bm, bj, ij\}$, is the donwnlink precoding/beamforming vector for cellular user $v$ in cell $u$. 
We use ${h}^{(\text{dc})}_{j,bm} \in \mathbb{C}$ ($L^{(\text{dc})}_{j,bm} \in \mathbb{R}_+$) to denote the channel (path-loss) from D2D transceiver $j$ to cellular user $m$ in cell $b$.
Further, $x^{(\text{d})}_{j} \in \mathbb{C}$ refers to the signals sent by D2D transceiver $j$, while $x^{(\text{c})}_{uv} \in \mathbb{C}, \, uv\in \{bm, bj, ij\},$ denotes the signal from BS $u$ intended to transmit to the cellular user $v$. 
In addition, we assume that all transmitted signals are normalized so that $\lvert x^{(\text{c})}_{uv} \rvert^2 = \lvert x^{(\text{d})}_j \rvert^2 = 1, \, u=1,\dots,B, \forall v \in \mathcal{C}_u, \, \forall j \in \mathcal{D}$. 
Further, $z_{bm}^{(\text{c})} \in \mathbb{C}$ refers to the thermal noise at cellular user $m$ in cell $b$, and it is distributed as $z_{bm}^{(\text{c})} \sim \mathcal{CN}(0, \sigma^2)$.

Consider a typical D2D link constituted by two D2D transceivers $n$ and $n'$ that can communicate simultaneously in both directions using the same channel. The received signal at D2D transceiver $n$ is given by 
\begin{align} \label{eq:signal:D2D} \small
	y^{(\text{d})}_{n} &= \sqrt{P_n} L_{n',n}^{(\text{dd})} {h}^{(\text{dd})}_{n',n} x_{n}^{(\text{d})}  \nonumber \\
	&+ \underbrace{ \sum\limits_{b=1}^{B} \sum\limits_{m\in\mathcal{C}_b} L^{(\text{cd})}_{b,n} \left(\mathbf{h}^{(\text{cd})}_{b,n} \right)^{\mathsf{H}} \mathbf{v}_{bm} x_{bm}^{(\text{c})}}_\textrm{Cellular-to-D2D interference} \nonumber \\
	&+ \underbrace{\sum\limits_{j \in \mathcal{D} \setminus \{n,n'\} } \sqrt{P_j} L_{j,n}^{(\text{dd})} {h}^{(\text{dd})}_{j,n} x_j^{(\text{d})} }_\textrm{Intra-D2D interference} + s_{n} + z_{n}^{(\text{d})}, 
\end{align} 
where $x_{u}^{(\text{d})} \in \mathbb{C}, \, u\in\{n,j\}$, denotes the signal sent by the D2D transceiver $u \in \mathcal{D}$, and $\mathbf{h}^{(\text{cd})}_{b,n} \in \mathbb{C}^A$ ($L^{(\text{cd})}_{b,n} \in \mathbb{R}_+$) denotes the vector channel (path-loss) from D2D transceiver $n$ to the BS $b$. 
In addition, ${h}^{(\text{dd})}_{j,n} \in \mathbb{C}$ ($L^{(\text{dd})}_{j,n} \in \mathbb{R}_+$) denotes the channel (path-loss) from D2D transceiver $j$ to the D2D transceiver $n$. 
Here, $s_n$ is the residual self-interference (SI) caused by imperfect cancellation of FD operation. We adopt an SI model in which the residual interference is reflected in the self-interference-to-power-ratio (SIPR) $\beta$ so that $\lvert s_n \rvert^2 = \beta P_n$ given by \cite{SImodelTong, FDAlouini} with $P_n$ being the instantaneous transmit power at the D2D transceiver $n$. $z_{n}^{(\text{d})} \in \mathbb{C}$ is the thermal noise at D2D transceiver $n$ and it is also distributed as $z_{n}^{(\text{d})} \sim \mathcal{CN}(0, \sigma^2), \, \forall n \in \mathcal{D}$. 
From (\ref{eq:signal:D2D}), the received signal at other D2D transceiver $n'$ can be obtained by replacing the index $n$ with $n'$ and vise versa, but we omit its derivation here for the sake of a concise presentation. 

This work assumes uniform linear array (ULA) employed at each BS. We model the channel vector $\mathbf{h}^{(\cdot)}$ from D2D/cellular users to BS as 
$ 
\mathbf{h}^{(\cdot)} = \left[\mathbf{A} \quad \mathbf{0}_{A\times A - P}\right] \times  \mathbf{\hat{h}}^{(\cdot)}, 
$
where $ \mathbf{\hat{h}}^{(\cdot)}$ denotes the fast fading channel vector \cite{mMIMO_Hoydis}. Each vector element is independent and identically distributed (i.i.d.) and follows the Rayleigh distribution with zero mean and unit variance, i.e., $\mathcal{CN}(0,1)$. Additionally, $\mathbf{0}_{A\times A - P}$ is the $A\times A - P$ zero matrix. Here, $P$ represents the fixed number of angular dimensions. The steering matrix $\mathbf{A} = \left[\mathbf{a}(\phi_1), \dots, \mathbf{a}(\phi_P)\right]  \in \mathbb{C}^{A \times P}$ is composed of the steering vector $\mathbf{a}(\phi)$ defined as 
$$ \small
\mathbf{a}(\phi) = \frac{1}{\sqrt{P}} \left[1, e^{-\text{i} 2\pi w \sin(\phi)} , \dots, e^{-\text{i} 2\pi w (A-1)\sin(\phi)}\right], 
$$
where $w$ is the antenna spacing in multiples of the wavelength and $\phi_p = -\pi /2 + (p-1)\pi/P, \, p = 1, \dots, P$, are uniformly distributed angles of transmission. 

Before processing further, to avoid cumbersome equations with many denotations in our subsequent derivations, we define the overall channel power gains as 
$g^{(\text{dc})}_{j,bm} \triangleq L^{(\text{dc})}_{j,bm} {h}^{(\text{dc})}_{j,bm},$ $g^{(\text{dd})}_{u,n} \triangleq L^{(\text{dd})}_{u,n} {h}^{(\text{dd})}_{u,n}, \, u\in \{n', j\},$ 
	$\left(\mathbf{g}^{(\text{cd})}_{b,n}\right)^{\mathsf{H}}\triangleq L^{(\text{cd})}_{b,n} \left(\mathbf{h}^{(\text{cd})}_{b,n} \right)^{\mathsf{H}}, $ and
	$\left(\mathbf{g}^{(\text{cc})}_{l,bm}\right)^{\mathsf{H}} \triangleq L^{(\text{cc})}_{l,bm} \left(\mathbf{h}^{(\text{cc})}_{l,bm} \right)^{\mathsf{H}},\,  l\in \{b,i\}. $


\subsection{Problem Formulation}
Given the system model, we now define the desired performance metrics, including signal-to-interference-and-noise-ratio (SINR) and sum-rate of cellular/D2D links, and correspondingly formulate the optimization problem. 
From the received signal in (\ref{eq:signal:cell}), the SINR at the cellular receiver $m$ in cell $b$ can be written as 
\begin{align} \label{eq:Cell:SINR} \small
	&\text{SINR}_{bm}^{(\text{c})} = \frac{\left\lvert \left(\mathbf{g}^{(\text{cc})}_{b,bm} \right)^{\mathsf{H}} \mathbf{v}_{bm} \right\rvert^2}{I^{(\text{c})}_{bm} +\sigma^2},
\end{align}
where the term $I^{(\text{c})}_{bm}$ represents the aggregate interference power caused by both cellular and D2D transmission, and it is given by
\begin{align} \label{eq:Inter:Cell} \small
	I^{(\text{c})}_{bm} = \sum\limits_{j\in \mathcal{C}_b\setminus \{m\}} &\left \lvert \left(\mathbf{g}^{(\text{cc})}_{b,bm}\right)^{\mathsf{H}}  \mathbf{v}_{bj} \right\rvert^2 + \sum\limits_{i \neq b } \sum\limits_{j\in \mathcal{C}_i}  \left \lvert \left(\mathbf{g}^{(\text{cc})}_{i,bm}\right)^{\mathsf{H}}  \mathbf{v}_{ij} \right\rvert^2  \nonumber \\
	&+ \sum\limits_{j \in \mathcal{D}}  P_{j} \left\lvert {g}^{(\text{dc})}_{j,bm} \right\rvert^2.
\end{align}
Treating the interference as noise, the achievable rate of cellular link $m$ in cell $b$ can be computed by invoking the Shannon's capacity formula as
\begin{align} \label{eq:Cell:Rate}
	R_{bm }^{(\text{c})} = \log_2\left(1+ \text{SINR}_{bm}^{(\text{c})} \right). 
\end{align} 

Meanwhile, for a FD  D2D link constituted by D2D transceivers $n$ and $n'$, the received SINR of D2D the transceiver $n$ can be written from (\ref{eq:signal:D2D}) as
\begin{align} \label{eq:D2D:SINR} \small
	&\text{SINR}_{n}^{(\text{d})} = \frac{P_n \left\lvert g^{(\text{dd})}_{n',n} \right\rvert^2 }{I_n^{(\text{d})} + \beta P_n +\sigma^2},
\end{align}
where 
\begin{align*}  \small 
	&I_n^{(\text{d})} \triangleq \sum\limits_{b=1}^B \sum\limits_{m\in \mathcal{C}_b} \left \lvert \left(\mathbf{g}^{(\text{cd})}_{b,n}\right)^{\mathsf{H}} \mathbf{v}_{bm} \right\rvert^2 + \sum\limits_{j \in \mathcal{D} \setminus \{n, n'\}}P_j  \left\lvert g^{(\text{dd})}_{j,n} \right\rvert^2 
\end{align*}
denotes the aggregate interference power caused by other D2D and cellular transmission. 
Likewise, the received SINR at D2D the transceiver $n'$, denoted as $\text{SINR}_{n'}^{(\text{d})}$, can be obtained by replacing the index $n$ with $n'$ in $\text{SINR}_{n}^{(\text{d})}$ and $I_n^{(\text{d})}$ and vise versa. 
As a result, the achievable rate of corresponding FD D2D link is given by
\begin{align} \label{eq:D2D:Rate} 
	R_{n}^{(\text{d})} = \log_2\left(1+ \text{SINR}_{n}^{(\text{d})}\right) + \log_2\left(1+ \text{SINR}_{n'}^{(\text{d})}\right).
\end{align}

This work uses the network sum-rate of both cellular and D2D links over a cell as the optimization objective under the constraints on the target (minimum required) SINR at both cellular and D2D transmission and maximum transmit power of D2D transmitters and BSs, denoted as $P_d$ and $P_c$, respectively. 
Without generality, we can assume that D2D (cellular) users have equal target SINR $\gamma_d\,(\gamma_c)$. 
The optimization problem therefore can be mathematically formulated as \vspace{-1ex}
\begin{subequations} \label{eq:Opt:Prob}
	\begin{align} \small
		\max\limits_{\mathbf{p}, \mathbf{V}} & \quad R \triangleq  \frac{1}{B}\sum\limits_{b=1}^B
		\sum\limits_{k \in \mathcal{C}_b} R_{bm }^{(\text{c})} +   \frac{1}{B} \sum\limits_{n \in \mathcal{D}} R_{n }^{(\text{d})} \\
		\text{s.t.}& \quad \text{SINR}_{bm}^{(\text{c})}  \ge \gamma_c,\,  b =1,\dots,B,\, \forall m\in \mathcal{C}_b,\\
		& \quad \text{SINR}_{n}^{(\text{d})} \ge \gamma_d,\, \forall n\in \mathcal{D}, \\
		& \quad \sum\limits_{m\in \mathcal{C}_b}  |\mathbf{v}_{bm}|^2 \le P_c, \,  b =1,\dots,B,\\
		& \quad P_n \le P_d, \, \forall n \in \mathcal{D},
	\end{align}
\end{subequations}
where $\mathbf{V}$ refers to the beamformer collection $\{\mathbf{v}_{bm}\}$ and $\mathbf{p}$ is the D2D transmit power vector.

\section{Joint Beamforming and Power Allocation Algorithm} \label{Ch6:sec-DCAl}
In this section, we focus on addressing the optimal solution of (\ref{eq:Opt:Prob}a-e). The problem (\ref{eq:Opt:Prob}a-e) involves a continuous nonconvex optimization, and it is not possible to directly obtain the globally optimal solution. To overcome this issue, we propose to apply fractional programming (FP) tools \cite{FPWu} to transform (\ref{eq:Opt:Prob}a-e) into a sequence of convex problems in which each problem can be solved effectively by standard convex optimization techniques. More specifically, the FP-based approach exploits the fractional structure of the objective function so that we can develop a sequential convex programming algorithm that approximately locates the globally optimal point with a low complexity. Such an approach is possible when the objective function exhibits a sum-ratio form of $\sum_n f\left({A_n(\mathbf{x})}/{B_n(\mathbf{x})}\right)$. Additionally, $f(\cdot)$ is a nondecreasing and concave function, 
while the functions $A_n(\mathbf{x}):\mathbb{C}^u \to \mathbb{R}$ and $B_n(\mathbf{x}): \mathbb{C}^u \to \mathbb{R}, \, u\ge 1,$ are convex and concave w.r.t. $\mathbf{x}$, respectively. 

Let first consider the case in which the numerator function $A_n(\cdot)$ can be represented by a quadratic form $A_n(\mathbf{x}) = a_n^2(\mathbf{x})$ with $a_n(\mathbf{x}): \mathbb{C}^u \to \mathbb{R}$ being a multidimensional and real-value function.
By adopting the quadratic transform of $A_n/B_n$ to $2q_n a_n - q_n^2/B_n$ \cite{FPWu}, the optimization problem  \vspace{-0.5ex}
\begin{align} \label{OrgProb} \small
	\max\limits_{\mathbf{x}} & \quad \sum_n f\left({A_n(\mathbf{x})}/{B_n(\mathbf{x})}\right)\\
	\text{s.t.} & \quad \mathbf{x}\in \mathcal{X}, \nonumber
\end{align}
where $\mathcal{X}$ represents the nonempty convex set of constraints, is equivalent to \cite{FPWu} \vspace{-0.5ex}
\begin{align} \label{FRProb} \small
	\max\limits_{\mathbf{x}, q_n} & \quad \sum_n f\left(2q_n a_n(\mathbf{x}) - q_n^2/B_n(\mathbf{x})\right)\\
	\text{s.t.} & \quad \mathbf{x}\in \mathcal{X}. \nonumber 
\end{align}
In (\ref{FRProb}), $q_n  \in \mathbb{R}$ refers to an auxiliary variable, and it is optimized when $\mathbf{x}$ is held fixed as $q_n^{\star} =  a_n(\mathbf{x})/B_n$.

For alternative cases in which the numerator of $f$ is represented by an expanded multiplication $A_n(\mathbf{x}) = a_n(\mathbf{x}) \bar{a}_n(\mathbf{x})$ with $a_n(\mathbf{x}): \mathbb{C}^u\to \mathbb{C}$ being a multidimensional and complex-valued function, the problem (\ref{OrgProb}) is now equivalent to \cite{FPWu} \vspace{-0.5ex}
\begin{align} \label{FRProbComplex} \small
	\max\limits_{\mathbf{x}, q_n} & \quad \sum_n f\left(2\text{Re}\left(\bar{q}_n a_n(\mathbf{x})\right)  - \bar{q}_n B_n(\mathbf{x}) q_n\right)\\
	\text{s.t.} & \quad \mathbf{x}\in \mathcal{X}, \nonumber 
\end{align}
where the auxiliary variables $q_n \in \mathbb{C}$ are also optimized at $q_n^{\star} = a_n(\mathbf{x})/B_n(\mathbf{x}).$ Here, $\bar{q}_n$ is the complex conjugate of $q_n$. 

The argument function of each outer function $f$ is now convex w.r.t. $\mathbf{x}$ and $q_n$ and the transformed optimization problems (\ref{FRProb}) and (\ref{FRProbComplex}) become convex. As a result, the global optimal solution can be achieved by solving a sequence of convex optimization subproblems that find the optimal $\mathbf{x}$ and $q_n$ in an iterative fashion. 

Based on the transformed convex problems (\ref{FRProb}) and (\ref{FRProbComplex}), we now focus on resolving the joint beamforming and power allocation problem (\ref{eq:Opt:Prob}). 
Although the argument in each cellular/D2D link rate component of the objective function in (\ref{eq:Opt:Prob}) is not in a direct ratio form (i.e., $1 + \text{SINR}$), the SINR terms of such objective function are in fractional form.  
Since the logarithm function $\log_2(\cdot)$ is nondecreasing and concave and each SINR term resides in the logarithm function, the quadratic transforms can be directly applied to the SINR terms. The detailed procedure is provided as follows.

Applying the quadratic transform to each SINR term and following (\ref{FRProb}) and (\ref{FRProbComplex}), the optimization problem (\ref{eq:Opt:Prob}) can be transformed to 
\begin{align} \label{eq:Opt:Prob:FR} \small
	\max\limits_{\mathbf{p}, \mathbf{V}, \{q_{bm}^{(\text{c})}\}, \{ q^{(\text{d})}_{n}\}} & \,\,\, f_{\text{FR}} = \frac{1}{B}\sum\limits_{b=1}^B
	\sum\limits_{m \in\mathcal{C}_b} f_{bm }^{(\text{c})} +  \frac{1}{B}\sum\limits_{n \in \mathcal{D}} f_{n}^{(\text{d})}. \\
	\text{s.t.}& \qquad (\ref{eq:Opt:Prob}b) - (\ref{eq:Opt:Prob}e). \nonumber 
\end{align}
Here,  $\{q_{bm}^{(\text{c})}\}$ and $\{ q^{(\text{d})}_{n}\}$ refer to the collections of auxiliary variables corresponding to the cellular and D2D sum-rates, respectively. 
In the objective function of (\ref{eq:Opt:Prob:FR}), using $\small \left\lvert \left(\mathbf{g}^{(\text{cc})}_{b,bm} \right)^{\mathsf{H}}  \mathbf{v}_{bm} \right\rvert^2 = \left(\mathbf{g}^{(\text{cc})}_{b,bm} \right)^{\mathsf{H}}  \mathbf{v}_{bm} \left( \mathbf{v}_{bm} \right)^{\mathsf{H}} \mathbf{g}^{(\text{cc})}_{b,bm} $,  the component $f_{bm}^{(\text{c})}$ can be expressed from (\ref{FRProbComplex}) as follows
\begin{align} \label{eq:obj:Cell:FR} \small
	f_{bm }^{(\text{c})}  = \log_2 &\left( 1 + 2 \text{Re}\left(2\bar{q}_{bm }^{(\text{c})}\left(\mathbf{g}^{(\text{cc})}_{b,bm} \right)^{\mathsf{H}}  \mathbf{v}_{bm}\right)  \right. \nonumber \\
	& \left. - \bar{q}_{bm }^{(\text{c})}\left(I_{bm}^{(\text{c})} +\sigma^2 \right)q_{bm }^{(\text{c})}\right). 
\end{align}
Meanwhile, the component $f_{n}^{(\text{d})}$ is provided from (\ref{FRProb}) by
\begin{align} \label{eq:obj:D2D:FR} \small
	f_{n}^{(\text{d})} = \log_2&\left( 1 + 2 q_n^{(\text{d})} \sqrt{P_n} \left\lvert g^{(\text{dd})}_{n',n} \right\rvert \right. \nonumber \\
	& \left. - 2 \left(q_n^{(\text{d})}\right)^2 \left(I_n^{(\text{d})} + \beta P_n+ \sigma^2\right) \right). 
\end{align}
From (\ref{eq:obj:Cell:FR}) and (\ref{eq:obj:D2D:FR}), we observe that, as the logarithmic function $\log_2(\cdot)$ is nondecreasing and concave,
the optimization problem (\ref{eq:Opt:Prob:FR}) is a convex problem of $\mathbf{V}$ and $\mathbf{p}$ when the auxiliary variables $\{q_{bm}^{(\text{c})}\}$ and $\{q_{n}^{(\text{d})}\}$ are held fixed.
It follows that, the optimal $q_{bm}^{(\text{c})}$ and $q_n^{(\text{d})}$ for fixed $\mathbf{v}_{bm}$ and $P_n$ equals
\begin{align} \label{eq:q:Cell} \small
	q_{bm}^{(\text{c})} &= {\left(\mathbf{g}^{(\text{cc})}_{b,bm} \right)^{\mathsf{H}}  \mathbf{v}_{bm}} \left({I_{bm}^{(\text{c})} + \sigma^2}\right)^{-1}, \\ 
	q_n^{(\text{d})}& = {\sqrt{P_n} \left\lvert g^{(\text{dd})}_{n',n} \right\rvert} \left({I_n^{(\text{d})} + \beta P_n + \sigma^2}\right)^{-1}. \label{eq:q:D2D} 
\end{align}

The convex optimization problem (\ref{eq:Opt:Prob:FR}) allows to develop an iterative algorithm in order to solve (\ref{eq:Opt:Prob}) as follows. The iterative algorithm generates a sequence $\mathbf{V}$ and $\mathbf{p}$ to improve the optimal solutions. From the first feasible solution of $\mathbf{V}$ and $\mathbf{p}$ that is randomly generated, at each iteration, 
we compute the optimal values of $\{q_{bm}^{(\text{c})}\}$ and $\{q_n^{(\text{d})}\}$ and 
subsequently locate the optimal solution of the convex program (\ref{eq:Opt:Prob:FR}), 
which can be solved effectively by using standard convex programming techniques. 
\emph{Because the constraint set $(\ref{eq:Opt:Prob}b) - (\ref{eq:Opt:Prob}e)$ is convex, the sequence $\mathbf{V}$ and $\mathbf{p}$ always converges \cite{FPWu}. }
We can set to stop the iterative algorithm when the objective function $f_{\text{FR}}$ converges, i.e., 
its absolute improvement is less than a desired (pre-selected) threshold $\epsilon$. 
For convenience, the iterative joint beamforming and power allocation algorithm based on FR programming is
summarized in Algorithm 2. 

\begin{algorithm} \label{C6:alg}
	\caption{Iterative FP-based algorithm}	
	\begin{algorithmic}[1]	
		\State  Initiate a feasible solution of $\mathbf{v}_{bm}$ and $\mathbf{p}$ and choose $\epsilon$.
		\Repeat
		\State Compute $\{q_{bm}^{(\text{c})}\}$ and $\{q_n^{(\text{d})}\}$ given by (\ref{eq:q:Cell}) and
		(\ref{eq:q:D2D}).
		\State Solve the convex program (\ref{eq:Opt:Prob:FR}) to obtain optimal $\mathbf{V}^{\star}$, $\mathbf{p}^{\star}$, and $f^{\star}_{\text{FR}}$. 	 	
		\Until{$f_{\text{FR}}$ converges, i.e., $\left|f_{\text{FR}} - f^{\star}_{\text{FR}} \right| \le \epsilon$}
	\end{algorithmic}
\end{algorithm}


\section{Illustrative Results} \label{sec-SimulResults}
In this section, numerical results are presented to compare the achievable network sum-rates (per cell) of both cellular and D2D links, provided by FP-based algorithm developed in previous section, for the FD D2D, HD D2D, and pure massive MIMO cellular systems. The behaviors of such sum-rates under the effect of SIC levels and numbers of active D2D links per cell are also illustrated.  

Our Monte Carlo simulations are performed as follows. 
We consider a multi-cell network consisting of three hexagonal-cells as shown in Fig. \ref{Fig:SystemModel}. 
The cellular users and D2D transceivers are dropped randomly within the cell region. Each D2D transceiver is located uniformly in the circle where the radius equals a fixed D2D link distance $r$ and the corresponding D2D transceiver is located at the origin. The channel fading are generated independently according to complex Gaussian distribution with unit variance. 
With regards to the steering matrix $\mathbf{A}$, we select the number of angular dimension $P=A/2$ and the antenna spacing $w=0.3$ provided in \cite{mMIMO_Hoydis}.
In FP-based algorithm, the improvement threshold of iterative algorithm is chosen as $\epsilon$. Unless stated otherwise, the network parameters are provided in Table \ref{Table:Para}. 
 
\begin{table}[htb!] 
	\caption{Simulation Parameters}	
	\centering %
	\begin{tabular}{|c | c|}				
		\hline			
		Cell area & $\pi 500^2 $ m$^2$ \\ \hline		
		Number of cells & 3 \\ \hline	
		Number of antennas at BS $A$ & $\{16, 32\}$ \\ \hline
		Number of angular dimension $P=A/2$ \cite{mMIMO_Hoydis} & $\{8, 16\}$ \\ \hline
		Antenna spacing $\omega$ \cite{mMIMO_Hoydis} & $0.3$ \\ \hline
		Number of cellular users per cell $M$ & $4$ \\ \hline		
		Number of D2D links per cell $N$ & [$10,\,40$]\\		 \hline
		D2D link distance $r$ \cite{D2DR12} & $50$ m \\ \hline  
		Path-loss exponent $\alpha$   \cite{LTER14} & $3.76$\\ \hline  
		Path-loss reference $C$ \cite{LTER14} & $-15.3$ dB  \\ \hline	
		Total BS Tx power $P_c$  &  $46$ dBm \\ \hline
		Max. D2D Tx power $P_d$ & $23$ dBm \\ \hline
		Carrier frequency & $2$ GHz \\ \hline	
		Channel bandwidth & $10$ MHz \\ \hline 
		Noise PSD & $-174$ dBm/Hz \\ \hline		
		Receiver noise figure & $9$ dB \\ \hline
		Self-interference-to-power-ratio $\beta$ &$-100\,$dB\\ \hline
		Target SINR range $\gamma_C = \gamma_D$  & $0$ dB \\ \hline	
		Improvement threshold $\epsilon$ & $10^{-5}$ dB \\ \hline
	\end{tabular}
	\label{Table:Para}
\end{table}
\begin{figure}[htb!]
	\centering
	.	\includegraphics[scale=0.50]{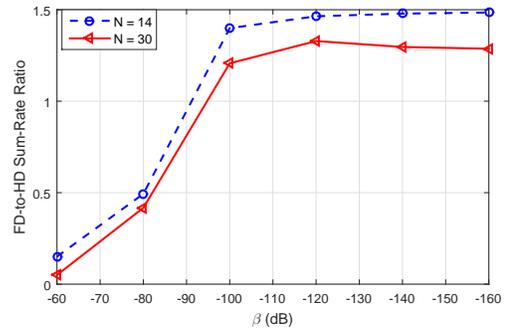}			
	\caption{FD-to-HD network sum-rate ratio versus SIPR.}
	\label{fig:RatioVsSIC}
\end{figure} 
For FD and HD D2D performance comparison, we first establish the FD-to-HD network sum-rate ratio and plot it versus SIPR $\beta$ in dB in Fig. \ref{fig:RatioVsSIC}. The numbers of cellular and D2D links per cell are chosen as $M=4$ and $N \in \{14, 30\}$ links/cell, respectively. 
In addition, the D2D link distance is $r=50\,$m, while the number of antennas equals $A=16$. 
The obtained results show that FD D2D can offer better sum-rate than HD D2D with SIPR $\beta \le -95$ dB. 
For $\beta \le -100\,$dB, the FD-to-HD D2D sum-rate ratios seem to be fixed around $1.45$ and $1.30$ for $N=14$ and $N=30$ links/cell, respectively. From the obtained results, we observe that the SIC level of $100\,$dB is sufficient to provide the best sum-rate ratio. 

%

\begin{figure} [htb!]
	\centering
	\includegraphics[scale=0.50]{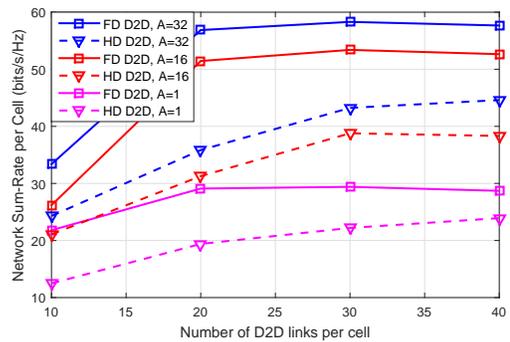}
	\caption{Achievable network sum-rate versus number of D2D links per cell.}
	\label{fig:RateVsNoD2D:Outdoor}
\end{figure} 
In Fig. \ref{fig:RateVsNoD2D:Outdoor}, we plot the achieved network sum-rate versus the number of D2D links over a cell under the consideration of FD and HD D2D. 
Furthermore, various numbers of antennas equipped at BSs are considered, i.e., $A \in\{1, 16, 32\}$, in order to demonstrate the advantage of ULA. Here, the case $A=1$ corresponds to the omni direction transmission at BSs. 
In the simulations for Fig. \ref{fig:RateVsNoD2D:Outdoor}, we assume the D2D link distance as $r=50\,$m, while the numbers of D2D links are from $N=10$ to $N=40$ links/cells. In addition, the SIC level is chosen as $\beta = -100\,$dB. Not surprisingly, increasing the number of equipped antennas at BSs allows to improve the sum-rate performances for both HD and FD D2D since narrower beams can be formed, thus reducing the interference caused by the cellular transmission. 
From Fig. \ref{fig:RateVsNoD2D:Outdoor}, it can be seen that the network sum-rate for both FD and HD D2D increases as the number of D2D links $N$ increases. However, such increase in sum-rate performance gets compressed when $N$ is beyond a certain threshold. For instance, with $N\ge 30$ links/cell and $A=16$ antennas ($A=32$ antennas), the FD-to-HD sum-rate ratio is fixed around $1.77$ ($1.31$). Beyond $N= 30$ links/cell, the increase in number of underlaid D2D links will not bring further sum-rate improvement.

\begin{figure} [htb!]
	\centering
	\includegraphics[scale=0.50]{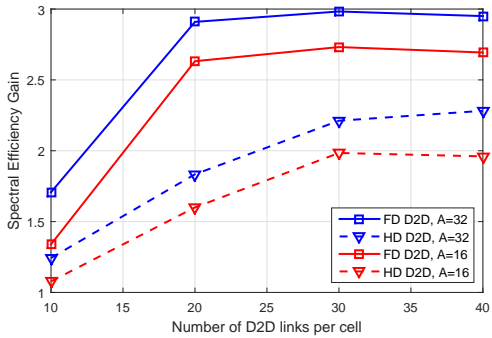}
	\caption{Spectral efficiency gain versus number of D2D links per cell.}
	\label{fig:SpecGain:Outdoor}
\end{figure} 
By defining the spectral efficiency gain as the ratio of the network sum-rate (provided in Fig. \ref{fig:RateVsNoD2D:Outdoor}) to the achievable sum-rate of only cellular links (operating alone without D2D links in the same environmental conditions), Fig. \ref{fig:SpecGain:Outdoor} plots the resulting spectral efficiency gain versus the number of active D2D link over a cell. The results in Fig. \ref{fig:SpecGain:Outdoor} indicate that, over the wide D2D link density range from $10$ to $40\,$links/cell, the spectral efficiency gain linearly increases with the D2D link number for the cases of FD and HD D2D to a sufficient high number of active D2D links. 
Additionally, it is beneficial to employ more antennas at BSs since the spectral efficiency gain is improved when the number of antennas increases from $A=16$ to $A=32$.
For instance, at D2D link number of $20\,$ links/cell and $A=16$ antennas ($A=32$ antennas), FD D2D offers a spectral efficiency gain of $2.61$ ($2.91$) as compared to pure cellular transmission. 
Meanwhile, HD D2D provides a spectral efficiency gain of $2.00$ ($2.21$) at the D2D link number of $30\,$ links/cell and $A=16$ antennas ($A=32$ antennas). Similar to the observation in Fig. \ref{fig:RateVsNoD2D:Outdoor},
for the number of active FD (HD) D2D links beyond $20$ $(30)$ links/cell, the increase in spectral efficiency gain will eventually get compressed. This result illustrates the  number of active D2D links should be chosen to optimize spectral efficiency gains offered by the underlaid D2D services.

\section{Conclusion} \label{sec-Conc}
In this paper, we have focused on multi-cell massive multiple-input-multiple-output (MIMO) cellular networks being underlaid by full-duplex (FD) device-to-device (D2D) transmission. Specifically, 
we proposed a joint beamforming and power allocation scheme that
maximizes the overall network throughput while protecting the D2D and
cellular link. To deal with the non-convexity of the problem,
a fractional programing (FP)-based method was developed to
transform the problem into a sequence of convex subproblems,
which can be solved efficiently. Simulation results
revealed that significant performance gains in term of spectral efficiency can be achieved in the considered FD network as compared to the half-duplex (HD) D2D counterpart and pure cellular transmission (in absent of D2D) with sufficient self-interference cancellation levels and number of underlaid D2D links.

\balance
\bibliographystyle{ieeetr}
\bibliography{IEEEabrv,Hung_IN}
\balance \balance

\end{document}